\documentclass[aps,prd,twocolumn,showpacs,amsmath,amssymb]{revtex4-1}
\usepackage{amsmath}
\usepackage{graphicx}
\usepackage{subfigure}
\usepackage{epstopdf}
\usepackage{color}
\usepackage{multirow}
\usepackage{setspace}
\usepackage{overpic}
\usepackage{amssymb}
\usepackage[bookmarksnumbered, pdfstartview=FitH,colorlinks,urlcolor=blue, citecolor=blue,linkcolor=blue] {hyperref}
\usepackage{lineno}
\usepackage{bm}
\usepackage{rotating}
\usepackage[utf8]{inputenc}

\let\oldequation\equation
\let\oldendequation\endequation

\renewenvironment{equation}
  {\linenomathNonumbers\oldequation}
  {\oldendequation\endlinenomath}

\begin{document}

\title{\bf \boldmath
Search for $D^0\to K^-\eta e^+\nu_e$, $D^+\to K_S^0 \eta e^+\nu_e$ and $D^+\to \eta\eta e^+\nu_e$ decays}

\author{M.~Ablikim$^{1}$, M.~N.~Achasov$^{4,c}$, P.~Adlarson$^{75}$, O.~Afedulidis$^{3}$, X.~C.~Ai$^{80}$, R.~Aliberti$^{35}$, A.~Amoroso$^{74A,74C}$, Q.~An$^{71,58,a}$, Y.~Bai$^{57}$, O.~Bakina$^{36}$
% ...
\\
\vspace{0.2cm}
(BESIII Collaboration)\\
\vspace{0.2cm} {\it
$^{1}$ Institute of High Energy Physics, Beijing 100049, People's Republic of China\\
% ...
$^{3}$ Bochum  Ruhr-University, D-44780 Bochum, Germany\\
$^{4}$ Budker Institute of Nuclear Physics SB RAS (BINP), Novosibirsk 630090, Russia\\
% ...
$^{35}$ Johannes Gutenberg University of Mainz, Johann-Joachim-Becher-Weg 45, D-55099 Mainz, Germany\\
$^{36}$ Joint Institute for Nuclear Research, 141980 Dubna, Moscow region, Russia\\
% ...
$^{57}$ Southeast University, Nanjing 211100, People's Republic of China\\
$^{58}$ State Key Laboratory of Particle Detection and Electronics, Beijing 100049, Hefei 230026, People's Republic of China\\
% ...
$^{71}$ University of Science and Technology of China, Hefei 230026, People's Republic of China\\
% ...
$^{74}$ University of Turin and INFN, (A)University of Turin, I-10125, Turin, Italy; (B)University of Eastern Piedmont, I-15121, Alessandria, Italy; (C)INFN, I-10125, Turin, Italy\\
$^{75}$ Uppsala University, Box 516, SE-75120 Uppsala, Sweden\\
% ...
$^{80}$ Zhengzhou University, Zhengzhou 450001, People's Republic of China\\
\vspace{0.2cm}
$^{a}$ Deceased\\
% ...
$^{c}$ Also at the Novosibirsk State University, Novosibirsk, 630090, Russia\\
% ...
}
}

\begin{abstract}
By analyzing $e^+e^-$ annihilation data corresponding to an integrated luminosity of 7.93 fb$^{-1}$, collected at the center-of-mass energy of 3.773 GeV  with the BESIII detector, we search for the semileptonic decays $D^0\to K^-\eta e^+\nu_e$,
$D^+\to K_S^0 \eta e^+\nu_e$ and $D^+\to \eta\eta e^+\nu_e$ for the first time.
We present evidence for $D^0\to K^-\eta e^+\nu_e$ with a significance of $3.3\sigma$.
The branching fraction of $D^0\to K^-\eta e^+\nu_e$ is measured to be $(0.84_{-0.34}^{+0.29}\pm0.22)\times 10^{-4}$. 
No significant signals are observed for the decays $D^+\to K_S^0 \eta e^+\nu_e$ and $D^+\to \eta\eta e^+\nu_e$ and we set the upper limits on their branching fractions. 
\end{abstract}

\maketitle

\oddsidemargin  -0.2cm
\evensidemargin -0.2cm

\section{Introduction}

\begin{figure}[htbp]
	\centering
	\includegraphics[width=0.45\linewidth]{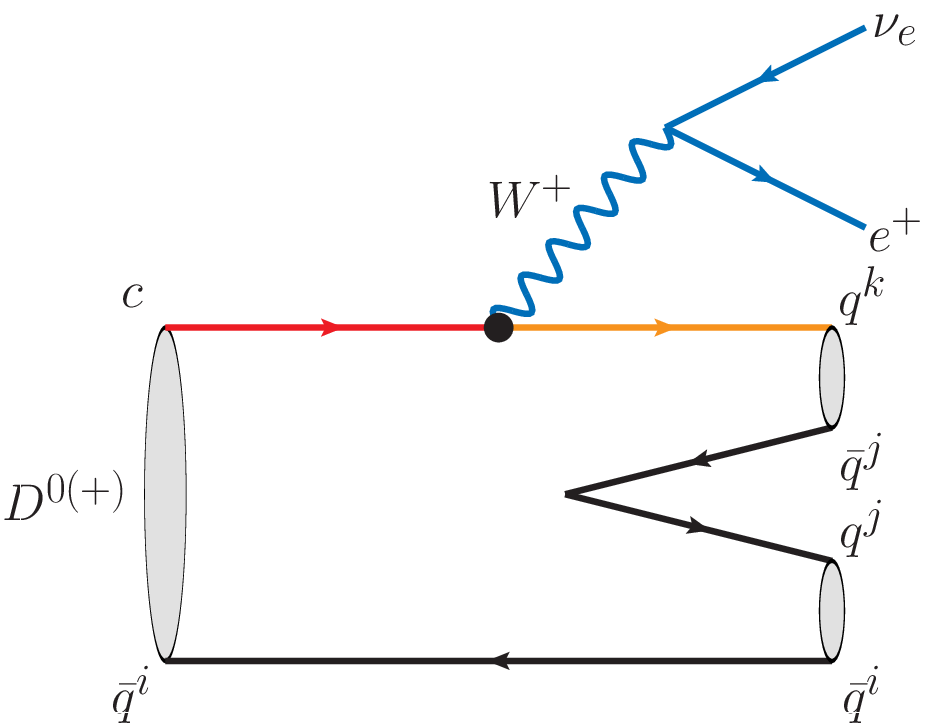}
	\includegraphics[width=0.45\linewidth]{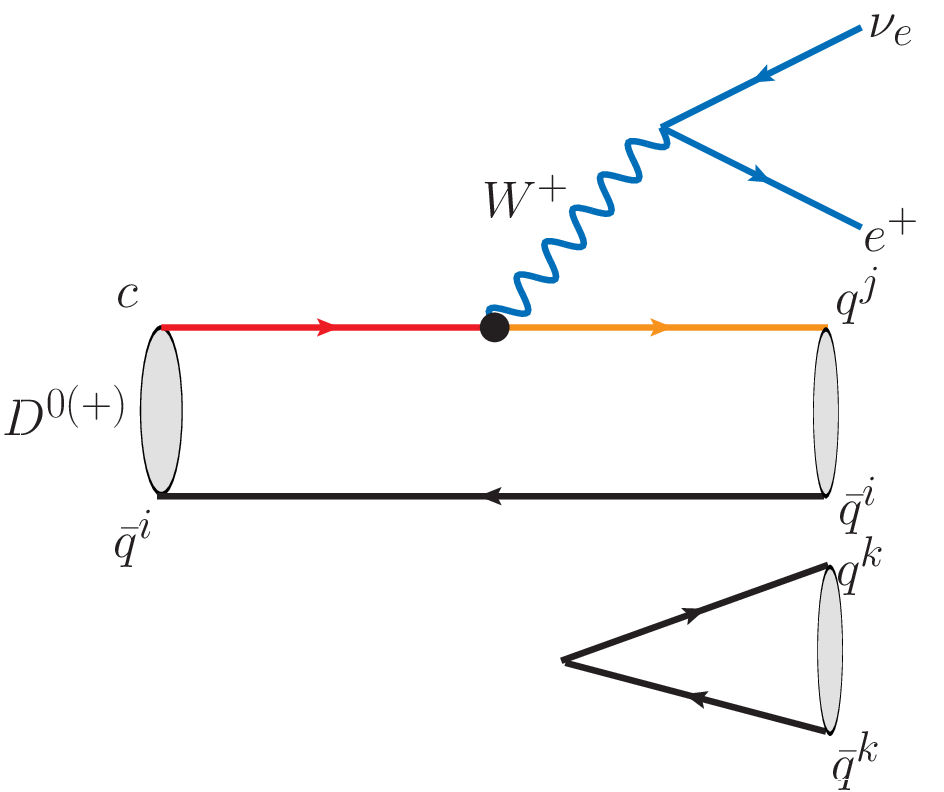}
	\caption{\footnotesize Two possible leading-order Feynman diagrams of the non-resonant decays $D^0\to K^-\eta e^+\nu_e$, $D^+\to K_S^0 \eta e^+\nu_e$ and $D^+\to \eta\eta e^+\nu_e$.}
	\label{fig:fey}
\end{figure}

Studies of the semileptonic $D$ decays are important to test the Standard Model (SM) and search for new physics beyond the SM~\cite{Ke:2023qzc}. Four-body semileptonic decays $D\to P_{1}P_{2}\ell^{+}\nu_{\ell}$ ($P$ denotes the pseudoscalar meson), mediated via $c\to (s,d)\ell^{+}\nu_{\ell}$ transitions, can receive contributions from light scalar or vector meson. Consequently, these decays provide a good laboratory for probing the internal structures of light hadrons~\cite{Wang:2009azc,Achasov:2012kk,Oset:2016lyh}. The experimental studies of the semileptonic decays $D\to P_{1}P_{2}\ell^{+}\nu_{\ell}$, except for the decays $D\to (V,S)\ell^{+}\nu_{\ell}$ with $(V,S)\to P_1P_2$ ($V,S$ denote a vector or scalar meson), are limited. The semileptonic decays $D^0\to K^-\eta e^+\nu_e$, $D^+\to K_S^0 \eta e^+\nu_e$ and $D^+\to \eta\eta e^+\nu_e$ proceed via the Feynman diagrams shown in Fig.~\ref{fig:fey}.

Throughout the paper, $D$ denotes a $D^0$ or $D^+$ meson. The branching fractions of $D^0\to K^-\eta e^+\nu_e$, $D^+\to K_S^0 \eta e^+\nu_e$ and $D^+\to \eta\eta e^+\nu_e$ decays are predicted~\cite{Wang:2022fbk} using SU(3) flavor symmetry theory, as summarized in Table~\ref{table:predict}. 
The branching fractions of $D\to V/S e^+\nu_{e}$ (where $V$ and $S$ denote $K_0^*(1430)$ and $f_0(1500)$, respectively) are predicted using the covariant light-front quark model~\cite{Cheng:2017pcq}.
No experimental study of these decays has been reported to date.

In this paper, we search for the semileptonic decays $D^0\to K^-\eta e^+\nu_e$,
$D^+\to K_S^0 \eta e^+\nu_e$ and $D^+\to \eta\eta e^+\nu_e$ for the first time, using an integrated luminosity of 7.93 fb$^{-1}$ of $e^+e^-$ collision data taken  with the BESIII detector at the center-of-mass energy of $\sqrt s =$ 3.773 GeV. Charge-conjugate modes are implied throughout this paper.

\begin{table}[htbp]
	\centering
	\caption{The  branching fractions of $D^0\to K^- \eta  e^+\nu_e$, $D^+\to \bar{K}^0 \eta e^+\nu_e$ and $D^+\to \eta \eta e^+\nu_e$ predicted using SU(3) flavor symmetry~\cite{Wang:2022fbk}.}
	\label{table:predict}
	\scalebox{0.82}{
		\begin{tabular}{lccc}
			\hline\hline
			Decay  & $D^{0}\to K^{-}\eta e^{+}\nu_{e}$ & $D^{+}\to \bar{K}^{0}\eta e^{+}\nu_{e}$ & $D^{+}\to\eta \eta e^{+}\nu_{e}$ \\ \hline
			Value ($\times10^{-6}$) & $3.51\pm 3.51$   & $8.9\pm 8.9$           & $3.16\pm 2.26$   \\
			\hline\hline
		\end{tabular}%
	}
\end{table}

\section{BESIII detector and Monte Carlo simulation}

The BESIII detector~\cite{BESIII} records symmetric $e^+e^-$ collisions 
provided by the BEPCII storage ring~\cite{BEPCII} in the center-of-mass energy range from 1.85 to  4.95~GeV, with a peak luminosity of $1 \times 10^{33}\;\text{cm}^{-2}\text{s}^{-1}$ 
achieved at $\sqrt{s} = 3.773\;\text{GeV}$. 
BESIII has collected large data samples in this energy region. The cylindrical core of the BESIII detector covers 93\% of the full solid angle and consists of a helium-based
 multilayer drift chamber~(MDC), a plastic scintillator time-of-flight
system~(TOF), and a CsI(Tl) electromagnetic calorimeter~(EMC),
which are all enclosed in a superconducting solenoidal magnet
providing a 1.0~T magnetic field.
The solenoid is supported by an
octagonal flux-return yoke with resistive plate counter muon
identification modules interleaved with steel. 
The charged-particle momentum resolution at $1~{\rm GeV}/c$ is
$0.5\%$, and the 
${\rm d}E/{\rm d}x$
resolution is $6\%$ for electrons
from Bhabha scattering. The EMC measures photon energies with a
resolution of $2.5\%$ ($5\%$) at $1$~GeV in the barrel (end cap)
region. The time resolution in the TOF barrel region is 68~ps, while
that in the end cap region was 110~ps. The end cap TOF
system was upgraded in 2015 using multigap resistive plate chamber
technology, providing a time resolution of 60~ps.  About 63\% of the data
used in this analysis benefits from this upgrade. 

Simulated data samples produced with a {\sc geant4}-based~\cite{geant4} Monte Carlo (MC) package, which includes the geometric description of the BESIII detector and the detector response, are used to determine detection efficiencies and to estimate backgrounds.
The simulation models the beam energy spread and initial state radiation (ISR) in the $e^+e^-$ annihilations with the generator {\sc kkmc}~\cite{kkmc1,kkmc2}.
The inclusive MC samples include the production of $D\bar{D}$ pairs (including quantum coherence for the neutral $D$ channels), the non-$D\bar{D}$ decays of the $\psi(3770)$, the ISR production of $J/\psi$ and $\psi(3686)$ states, and the continuum processes incorporated in {\sc kkmc}~\cite{kkmc1,kkmc2}.
All particle decays are modelled with {\sc evtgen}~\cite{evtgen1,evtgen2} using the branching fractions either taken from the Particle Data Group~\cite{pdg2022} when available, or otherwise estimated with {\sc lundcharm}~\cite{lundcharm}.
Final state radiation~(FSR) from charged final state particles is incorporated using the {\sc photos} package~\cite{photos}.
Because there is little information about the decays understudied, the signal decays $D\to P\eta e^+\nu_e$ are simulated, proceeding with and without sub-resonance, with a fraction of half for each. For the first case, they are generated as $D\to {\mathcal R} e^+\nu_e$ with ${\mathcal R}\to P\eta$ by using the ISGW2 model~\cite{isgw2}, where $P = K$, $K_{S}^{0}$, and $\eta$. For the second case, called phase-space (PHSP) model, they are generated in n-body decays, averaging over the spins of initial and final state particles.

\section{Measurement method}

The  $e^+e^-\to \psi(3770)\to D\bar D$ process provides an ideal platform to investigate semileptonic $D$ decays by using the double-tag~(DT) method~\cite{MARK-III:1985hbd,MARK-III:1987jsm}, benefiting from the fact that there is no additional particle accompanying the $D\bar D$ pair in the final state.
First, single-tag~(ST) $\bar D^0$ mesons are reconstructed by using the hadronic decay modes 
$\bar D^0\to K^+\pi^-$, $K^+\pi^-\pi^0$, and $K^+\pi^-\pi^-\pi^+$,
while ST $D^-$ mesons are reconstructed via the decays
$D^-\to K^{+}\pi^{-}\pi^{-}$,
$K^0_{S}\pi^{-}$, $K^{+}\pi^{-}\pi^{-}\pi^{0}$, $K^0_{S}\pi^{-}\pi^{0}$, $K^0_{S}\pi^{+}\pi^{-}\pi^{-}$,
and $K^{+}K^{-}\pi^{-}$.
Then, signal candidates are reconstructed from the remaining tracks and showers.  Candidates in which $D^0\to K^-\eta e^+\nu_e$, $D^+\to K_S^0 \eta e^+\nu_e$ or $D^+\to \eta\eta e^+\nu_e$ with $\bar D$ decays into a ST mode is called DT candidates.  
The branching fractions of the signal decays are determined by
\begin{equation}
\label{eq:bf1}
{\mathcal B}_{\rm SL} = \frac{N_{\rm DT}}{N^{\rm tot}_{\rm ST}\cdot\bar \varepsilon_{\rm SL} \cdot {\mathcal B}_{\rm sub}},
\end{equation}
where $N_{\rm ST}^{\rm tot}$ represents the total number of ST $\bar D$ mesons,  $N_{\rm DT}$ denotes the number of the DT events,
${\mathcal B}_{\rm sub}$ is ${\mathcal B}_{\eta\to \gamma\gamma}$, ${\mathcal B}_{K^0_S\to \pi^+\pi^-}\cdot {\mathcal B}_{\eta\to \gamma\gamma}$, and ${\mathcal B}^2_{\eta\to \gamma\gamma}$
for the decays $D^0\to K^-\eta e^+\nu_e$, $D^+\to K_S^0 \eta e^+\nu_e$ and $D^+\to \eta\eta e^+\nu_e$, respectively.
${\mathcal B}_{K^0_S\to \pi^+\pi^-}$ and ${\mathcal B}_{\eta\to \gamma\gamma}$ are the branching fractions of $K^0_S\to\pi^+\pi^-$ and $\eta\to \gamma\gamma$. Additionally, $\bar\varepsilon_{\rm SL}$ is the average efficiency of reconstructing the semi-leptonic signal decays.
The average signal efficiency, weighted over the ST modes $i$, is calculated as $\bar\varepsilon_{\rm SL}=\Sigma_i [(\varepsilon^i_{\rm DT}\cdot N^i_{\rm ST})/(\varepsilon^i_{\rm ST}\cdot N^{\rm tot}_{\rm ST})]$, where  $N^i_{\rm ST}$ is the ST yield of $\bar{D}\to i$, $\varepsilon^i_{\rm ST}$ is the detection efficiency of reconstructing $\bar{D}\to i$, and $\varepsilon^i_{\rm DT}$ is the detection efficiency of reconstructing the ST $\bar D$ and the signal decay simultaneously.
The product of the branching
fractions of the semileptonic $D$ decay is determined via
\begin{equation}
\label{eq:bf2}
{\mathcal B}_{\rm SL} = \frac{N_{\rm DT}}{N^{\rm tot}_{\rm ST}\cdot\bar \varepsilon_{\rm SL}  \cdot {\mathcal B}_{\rm sub}}.
\end{equation}

\section{Selection of ST $\bar D$ candidates}

All charged tracks used in this analysis are required to be within a polar angle ($\theta$) range of $|\cos\theta|<0.93$, where $\theta$ is defined with respect to the $z$-axis, which is the symmetry axis of the MDC. 
For good charged tracks not originating from $K^0_S$ decays, the distance of closest approach to the interaction point (IP) must be less than 10\,cm along the $z$-axis, $|V_{z}|$, and less than 1\,cm in the transverse plane, $|V_{xy}|$.
Charged tracks are identified by using the energy deposited in the MDC ($dE/dx$) and the flight time in the TOF; combined likelihoods for the pion and kaon hypotheses are computed separately. Kaon and pion candidates are required to satisfy $\mathcal{L}_{K} > \mathcal{L}_{\pi}$ and $\mathcal{L}_{\pi} > \mathcal{L}_{K}$, respectively.

Each $K_S^0$ candidate mesons is reconstructed from two oppositely charged tracks each satisfying $|V_{z}|<$ 20~cm. 
The two charged tracks are assigned as $\pi^+\pi^-$ without imposing further particle identification (PID) criteria.
They are constrained to originate from a common vertex, which is required to be away from the IP by a flight distance of at least twice the vertex resolution. The quality of the vertex fits (a primary vertex fit and a secondary vertex fit for $K^0_S$) is ensured by a requirement of $\chi^{2} < 100$. The invariant mass of the $\pi^+\pi^-$ pair is required to be within $(0.487,0.511)$~GeV/$c^2$.

Neutral pion candidates are reconstructed by the $\pi^0\to\gamma\gamma$ decays. Photon candidates are identified using showers in the EMC. The deposited energy of each shower must be more than 25~MeV in the barrel region ($|\cos \theta|< 0.80$) and more than 50~MeV in the end cap region ($0.86 <|\cos \theta|< 0.92$)~\cite{BESIII}.  
To exclude showers that originate from charged tracks,
the angle subtended at the IP by the EMC shower and the position of the closest charged track at the EMC
must be greater than 10 degrees. 
To suppress electronic noise and showers unrelated to the event, the difference between the EMC shower time and the event start time is required to be within [0, 700]\,ns.
For $\pi^0$ candidates, the invariant mass of the photon pair is required to be within $(0.115,\,0.150)$\,GeV$/c^{2}$. To improve the momentum resolution, a mass-constrained~(1-C) fit to the nominal $\pi^{0}$ mass~\cite{pdg2022}; a fit $\chi^2<50$ is also required. 

\begin{figure}[htbp]\centering
\includegraphics[width=1.0\linewidth]{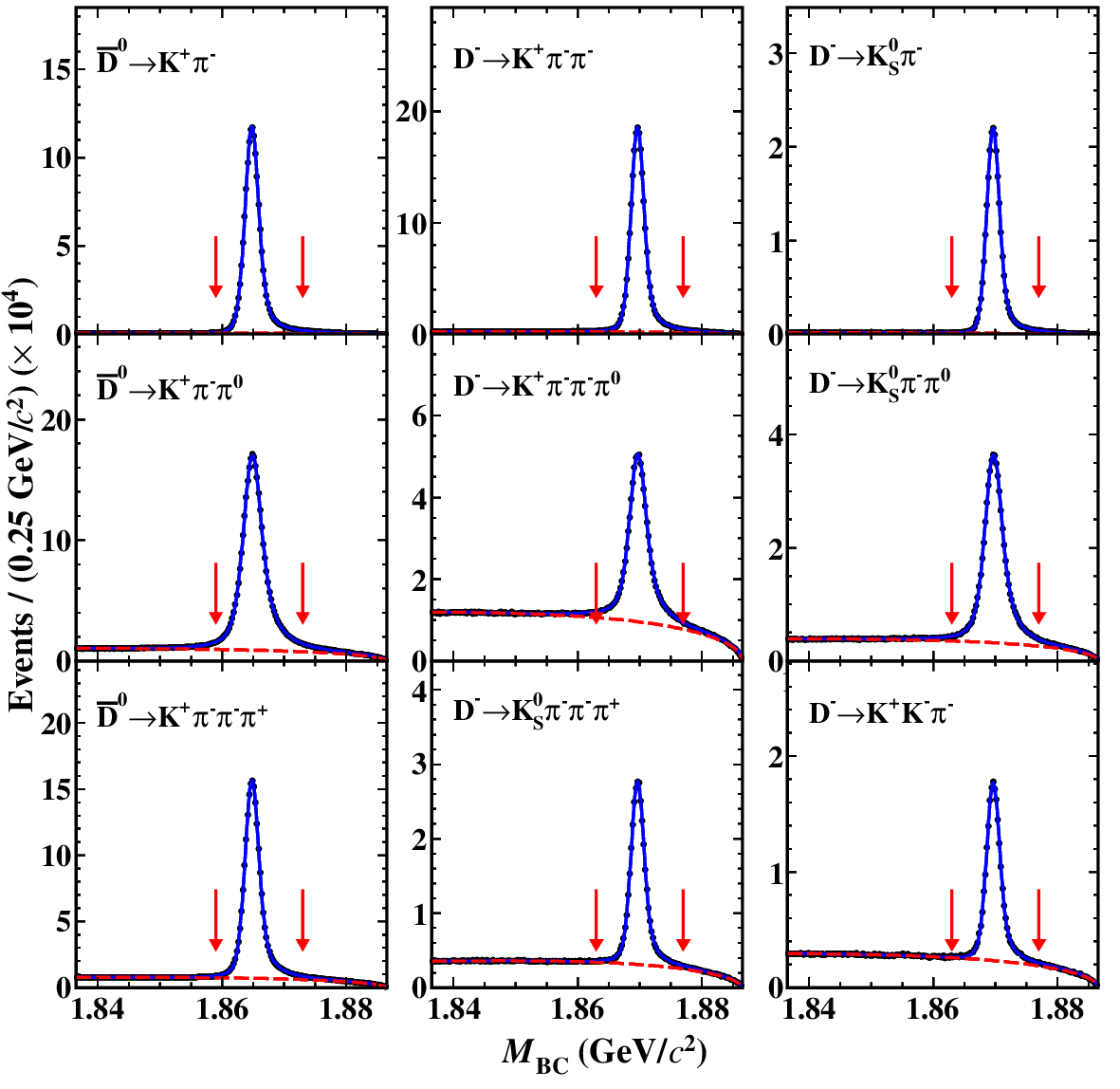}
\caption{The $M_{\rm BC}$ distributions for the different ST $\bar D$ decays with fit results overlaid. In each plot, the points with error bars are data, the red dashed curve is the background contribution, and the blue solid line shows the total fit.  Pairs of red arrows show the $M_{\rm BC}$ signal windows.}
\label{fig:datafit_Massbc}
\end{figure}

In the selection of $\bar D^0\to K^+\pi^-$ events, the backgrounds from cosmic rays and Bhabha events are rejected by using the same requirements described in Ref.~\cite{deltakpi}.
To separate the ST $\bar D$ mesons from combinatorial backgrounds, we define the energy difference $\Delta E\equiv E_{\bar D}-E_{\mathrm{beam}}$ and the beam-constrained mass $M_{\rm BC}\equiv\sqrt{E_{\mathrm{beam}}^{2}/c^{4}-|\vec{p}_{\bar D}|^{2}/c^{2}}$, where $E_{\mathrm{beam}}$ is the beam energy, and $E_{\bar D}$ and $\vec{p}_{\bar D}$ are the total energy and momentum of the ST $\bar D$ candidate in the $e^+e^-$ center-of-mass frame.
If there is more than one $\bar{D}$ candidate in a specific ST mode, the one with the minimum $|\Delta E|$ value is kept for further analysis.
The corresponding $|\Delta E|$ requirements for each ST modes are listed in Table~\ref{tab:st}. To extract the yield of ST $\bar D$ mesons for each mode, a fit is performed to the corresponding $M_{\rm BC}$ distribution. The signal is described by an MC-simulated shape convolved with a double-Gaussian function which compensates the resolution difference between data and MC simulation. The background is described by an ARGUS function~\cite{argus}. All fit parameters are left free in the fits.
Figure~\ref{fig:datafit_Massbc} shows the fits to the $M_{\rm BC}$ distributions for individual ST modes. Candidates within the $M_{\rm BC}$ signal region of $(1.864,1.876)$~GeV$/c^2$, shown as red arrows in Fig.~\ref{fig:datafit_Massbc}, are kept for further analysis. Combining the contributions from all ST modes, the total yields of ST $\bar D^0$ and $D^-$ mesons are obtained to be $N_{{\rm ST}, \bar{D}^0}^{\rm tot} = (6306.9\pm 2.8_{\rm stat})\times 10^3$ and $N_{{\rm ST}, D^-}^{\rm tot} = (4149.9\pm 2.4_{\rm stat})\times 10^3$, respectively.

\begin{table}[htbp]
	\centering
	\caption {\label{tab:st}The $\Delta E$ requirements, the obtained ST $\bar D^0(D^-)$ yields ($N^{i}_{\rm ST}$) in data and the ST efficiencies ($\varepsilon^{i}_{\rm ST}$). The efficiencies do not include the branching fractions of the $K^0_S$ and $\pi^0$ decays. The uncertainties are statistical only.}
	\label{table:styields}
	\scalebox{0.88}{
	\begin{tabular}{lccc}
		\hline\hline
		Tag mode 										& $\Delta E$~(GeV)   &  $N^{i}_{\rm ST}~(\times10^{3})$     	&  $\varepsilon^{i}_{\rm ST}~(\%)$  \\\hline
		$\bar{D}^{0} \to K^{+} \pi^{-}$ 				&  $(-0.027,0.027)$ & $1449.5 \pm 1.3$ 			& $64.95 \pm 0.01$ \\
		$\bar{D}^{0} \to K^{+} \pi^{-} \pi^{0}$ 		&  $(-0.062,0.049)$ & $2913.2 \pm 2.0$ 			& $35.52 \pm 0.01$ \\
		$\bar{D}^{0} \to K^{+} \pi^{+} \pi^{-} \pi^{-}$ &  $(-0.026,0.024)$ & $1944.2 \pm 1.6$ 			& $40.42 \pm 0.01$ \\
		\hline
		$D^{-} \to K^{+} \pi^{-} \pi^{-}$ 				&  $(-0.025,0.024)$ & $2164.0 \pm 1.6$ 			& $51.17 \pm 0.01$ \\
		$D^{-} \to K_{S}^{0} \pi^{-}$ 					&  $(-0.025,0.026)$ & $250.4 \pm 0.5$ 			& $50.63 \pm 0.02$ \\
		$D^{-} \to K^{+} \pi^{-} \pi^{-} \pi^{0}$ 		&  $(-0.057,0.046)$ & $689.0 \pm 1.2$ 			& $25.50 \pm 0.01$ \\
		$D^{-} \to K_{S}^{0} \pi^{-} \pi^{0}$ 			&  $(-0.062,0.049)$ & $558.5 \pm 0.9$ 			& $26.28 \pm 0.01$ \\
		$D^{-} \to K_{S}^{0} \pi^{-} \pi^{-} \pi^{+}$ 	&  $(-0.028,0.027)$ & $300.5 \pm 0.7$ 			& $28.97 \pm 0.01$ \\
		$D^{-} \to K^{+} K^{-} \pi^{-}$ 				&  $(-0.024,0.023)$ & $187.4 \pm 0.5$ 			& $41.06 \pm 0.02$ \\
		\hline\hline
	\end{tabular}
	}
\end{table}

\section{Selection of DT $D \bar D$ events}

In the presence of the ST $\bar D$ candidates,
signal decays are reconstructed from charged tracks and showers which have not been used in the ST selection.
It is required that there are exactly two, three, or one charged tracks reconstructed in $D^0\to K^-\eta e^+\nu_e$, $D^+\to K^0_S\eta e^+\nu_e$, or $D^+\to \eta \eta e^+\nu_e$ decays, respectively.

The positron candidate is required to have a charge opposite to that of the charm quark in the ST $\bar D$ meson.  
Information from $dE/dx$, TOF, and the EMC measurements is combined to create combined likelihoods under the positron, pion, and kaon hypotheses~($\mathcal{L}_e$, $\mathcal{L}_\pi$, and $\mathcal{L}_K$).  
The positron candidate is required to satisfy $\mathcal{L}_e>0.001$ and $\mathcal{L}_e/(\mathcal{L}_e+\mathcal{L}_\pi+\mathcal{L}_K)>0.8$. 
To reduce background from hadrons and muons, the positron candidate is further required to have a an $E/p$, the deposited energy in the EMC divided by the momentum measured in the MDC, between 0.8 and 1.1. 

The invariant mass of the $\eta$ candidate photon pair is required to be within $(0.505,\,0.575)$\,GeV$/c^{2}$. To improve the momentum resolution, a mass-constrained~(1-C) fit to the nominal $\eta$ mass~\cite{pdg2022} is imposed on the photon pair.  
The fit $\chi^2$ must be less than 50 and the four-momentum of the $\eta$ candidate returned by this kinematic fit is used for further analysis.
The selection criteria of photons, charged and neutral kaons are the same as those used in the ST selection.

Peaking backgrounds from hadronic $D$ decays with multiple pions in the final states are rejected by requiring that the invariant mass of the $P\eta e^+$ system, $M_{P\eta e^+}$, is less than 1.80~GeV/$c^2$. To suppress backgrounds with extra photon(s), we require that the energy of any extra photon, $E_{\rm extra}^{\gamma}$, is less than 0.25~GeV and there is no extra $\pi^0$ ($N_{\rm extra}^{\pi^0}=0$) in the candidate event. To reject the background from non-$\eta$ decays, such as $D^{0}\to K^{-}\pi^0 e^{+}\nu_{e}, K^-\pi^+\pi^0\pi^0$, 
in the $D^{0}\to K^{-}\eta e^{+}\nu_{e}$ analysis, we require the invariant mass of the $K^{-} \eta$ system to satisfy $M_{K^{-} \eta}>1.30~{\rm GeV}/c^2$.

The neutrino cannot be directly detected in the BESIII detector. Instead, information about the missing neutrino is inferred by the kinematic quantity
$U_{\mathrm{miss}}\equiv E_{\mathrm{miss}}-|\vec{p}_{\mathrm{miss}}|$.  Here, $E_{\mathrm{miss}}$ and $\vec{p}_{\mathrm{miss}}$ are the missing energy and momentum of the semileptonic candidate, respectively, calculated by $E_{\mathrm{miss}}\equiv E_{\mathrm{beam}}-\Sigma_j E_j$
and $\vec{p}_{\mathrm{miss}}\equiv\vec{p}_{D}-\Sigma_j \vec{p}_j$ in the $e^+e^-$ center-of-mass frame. The index $j$ sums over the $P$, $\eta$ and $e^+$ in the signal candidate, and $E_j$ and $\vec{p}_j$ are the energy and momentum of the $j$th particle, respectively.
To improve the $U_{\mathrm{miss}}$ resolution, the $D$ energy is constrained to the beam energy and $\vec{p}_{D}
\equiv -\hat{p}_{\bar D}\sqrt{E_{\mathrm{beam}}^{2}-m_{D}^{2}}$, where
$\hat{p}_{\bar D}$ is the unit vector in the direction of
the ST $\bar D$ momentum, and $m_{D}$ is the $D$ nominal
mass~\cite{pdg2022}. For correctly reconstructed signal events, $U_{\mathrm{miss}}$  peaks at zero.

The detection efficiencies $\bar \varepsilon_{\rm SL}$ for $D^0\to K^-\eta e^+\nu_e$, $D^+\to K^0_S\eta e^+\nu_e$ and $D^+\to \eta \eta e^+\nu_e$ are
($5.29\pm0.05$)\%, ($7.79\pm0.05$)\% and ($7.52\pm0.05$)\%, respectively.

\begin{figure*}[htbp]
	\includegraphics[width=0.328\linewidth]{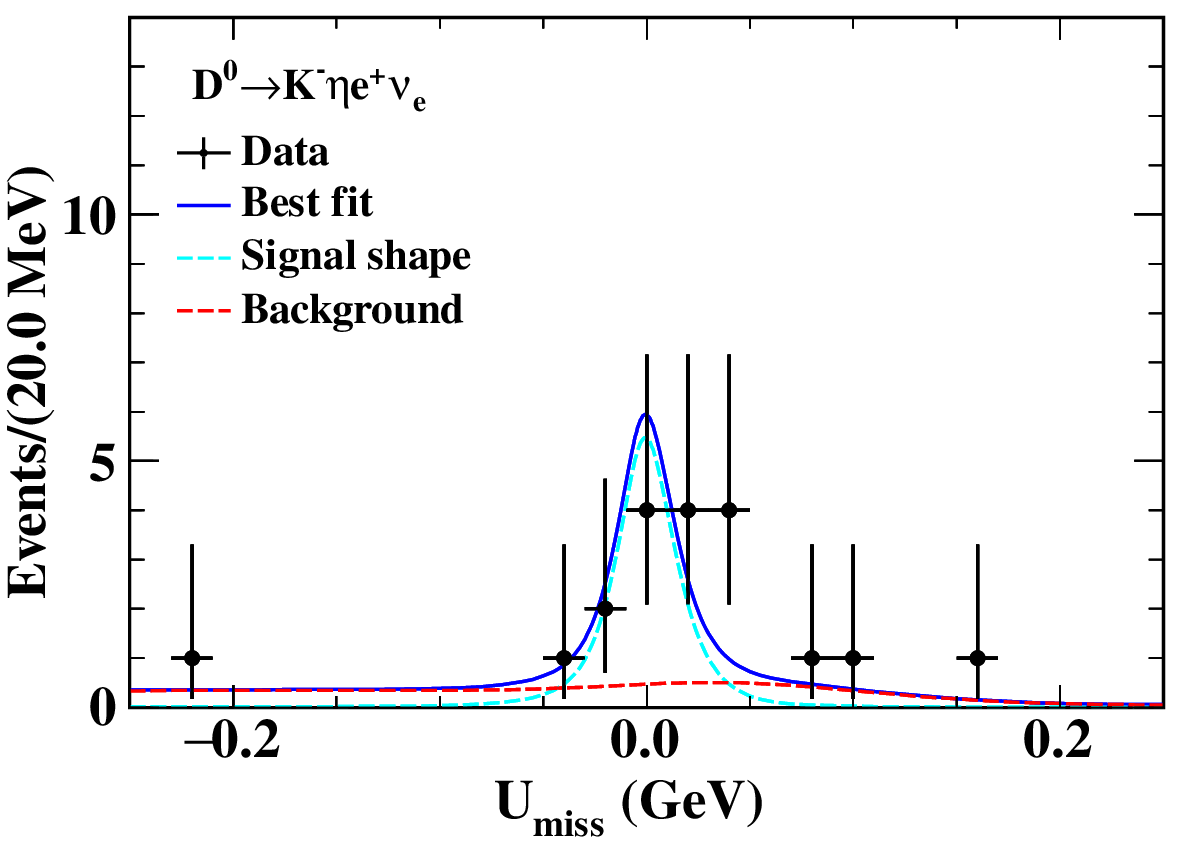}
	\includegraphics[width=0.328\linewidth]{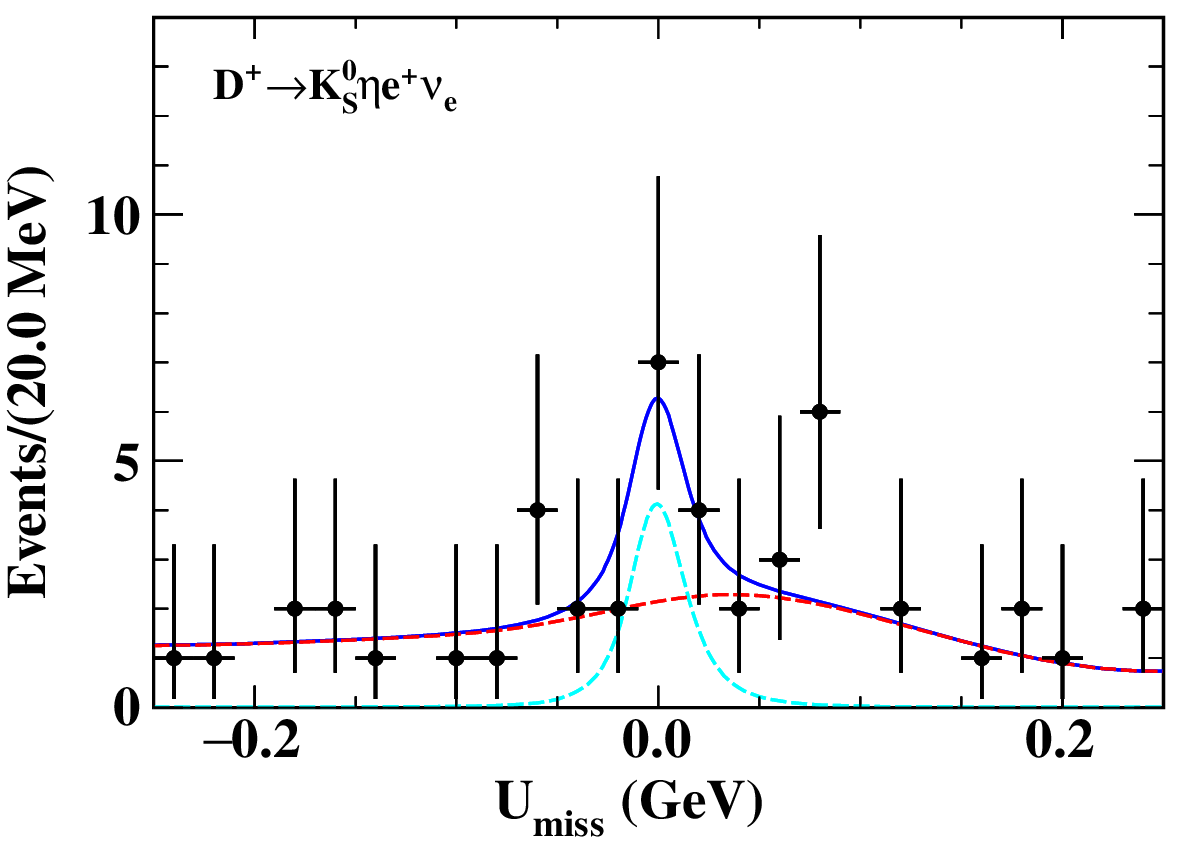}
	\includegraphics[width=0.328\linewidth]{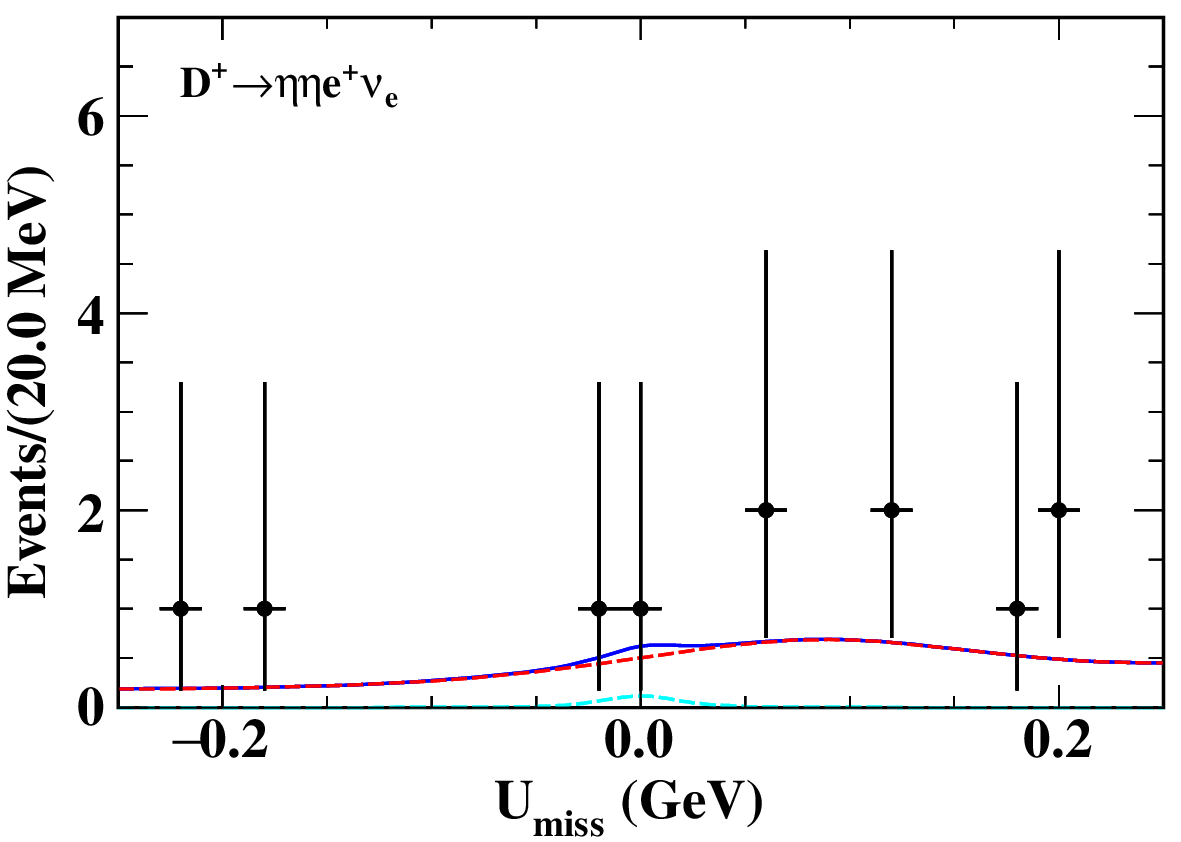}
	\caption{The $U_{\rm miss}$ distributions of the
		$D^0\to K^-\eta e^+\nu_e$, $D^+\to K_S^0 \eta e^+\nu_e$ and $D^+\to \eta\eta e^+\nu_e$ candidate events with fit results overlaid.
		The points with error bars are data, the light blue dashed curve is the signal, the red dashed curve is the background contribution, and the blue solid curve shows the total fit.}
	\label{fig:fit_Umistry1}
\end{figure*}

Figure~\ref{fig:fit_Umistry1} shows the $U_{\rm miss}$ distributions of the accepted candidate events. To extract the signal yields of each signal decay,
we perform an unbinned maximum likelihood fit on the corresponding $U_{\rm miss}$ distribution. In the fit, the signal and background components are characterized by the simulated shapes obtained from the signal MC events and the inclusive MC sample, respectively, with signal and background yields floating.  

\begin{figure*}[htbp]
	\includegraphics[width=0.48\linewidth]{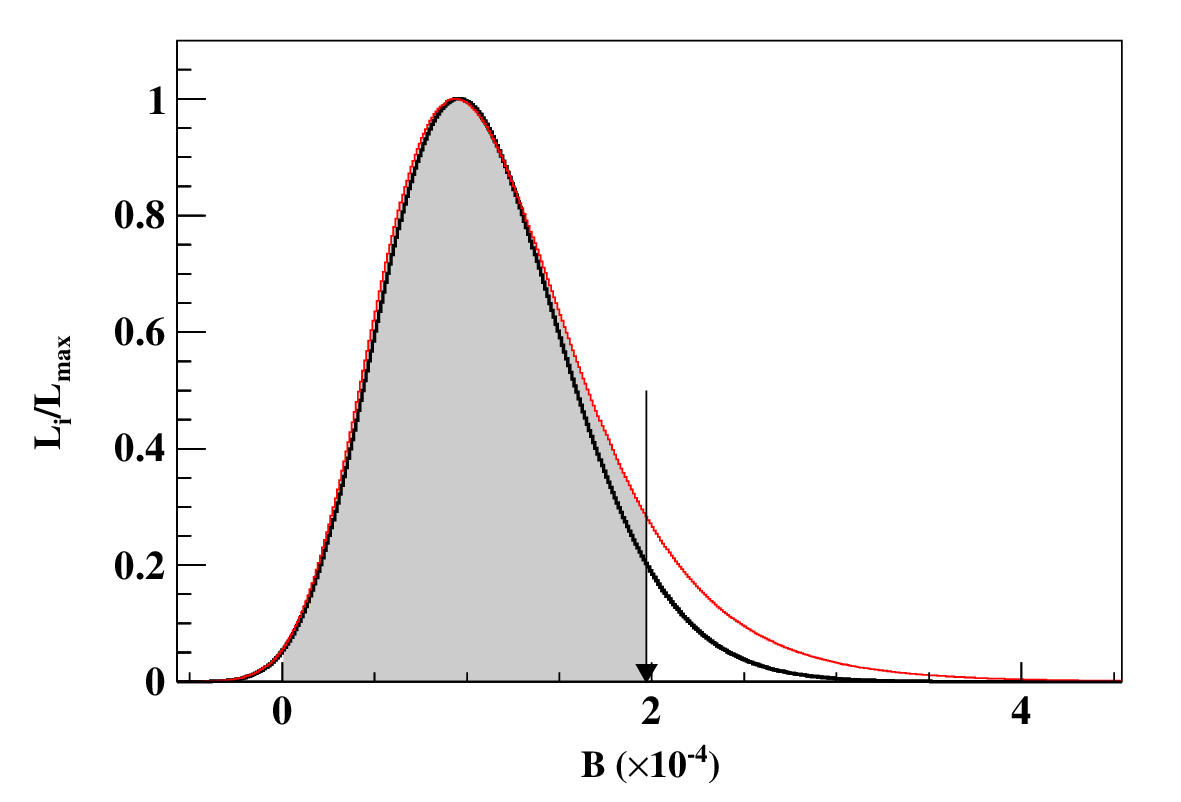}
	\includegraphics[width=0.48\linewidth]{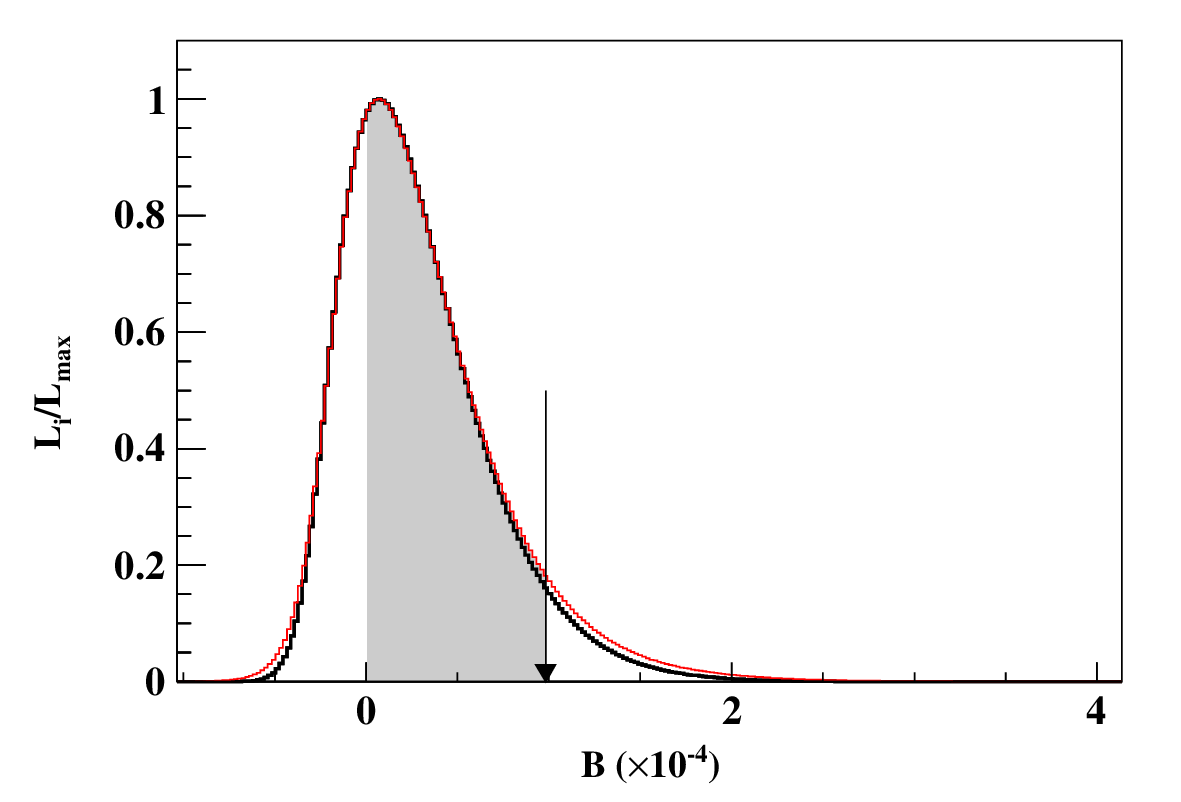}
	\caption{The normalized likelihood distributions versus the product of branching fractions of
		$D^+\to K_S^0 \eta e^+\nu_e$ (left) and $D^+\to \eta\eta e^+\nu_e$ (right) under the fully resonant assumption. The red solid (black dotted) curves indicate results including (not including) systematic uncertainties. The black arrows show the upper limits on the branching fractions at the 90\% confidence level.}
	\label{fig:prob}
\end{figure*}

Evidence for the decay $D^0\to K^-\eta e^+\nu_e$ is found with a statistical significance of $3.3\sigma$. The branching fraction of the decay $D^0\to K^-\eta e^+\nu_e$ is determined to be $(0.68_{-0.23}^{+0.27})\times 10^{-4}$.
As no significant signals are observed for the $D^+\to K^0_S\eta e^+\nu_e$ and $D^+\to \eta \eta e^+\nu_e$ decays,  upper limits on their branching fractions are set at the 90\% confidence level (C.L). The blue dotted curves in Fig.~\ref{fig:prob} show the normalized likelihood distributions versus the corresponding products of branching fractions.

\section{Systematic uncertainty}

With the DT method, many systematic uncertainties associated with the ST selection cancel out in the branching fraction measurements.
The sources of the systematic uncertainties in the measurements of the product of branching fractions are categorized into multiplicative and additive systematic uncertainties for the decays $D^+\to K_S^0\eta e^+\nu_{e}$ and $D^+\to\eta\eta e^+\nu_{e}$. However, the sources of the systematic uncertainties are not classified for the $D^0\to K^-\eta e^+\nu_e$ decay.
Multiplicative systematic uncertainties include sources uncertainty related to various efficiencies and are assigned relative to the measured branching fractions.
The uncertainty associated with the ST yield $N_{\rm ST}^{\rm tot}$ is estimated to be~0.1\% due to the fit to the $M_{\rm BC}$ distributions, which is studied by varying the signal and background shapes. The uncertainties of the quoted branching fractions of $K^0_S \to\pi^+\pi^-$ and $\eta\to \gamma\gamma$ are 0.07\% and 0.5\%~\cite{pdg2022}, respectively.
The uncertainties from the tracking and PID of $e^\pm$ are studied with a control sample of $e^+e^-\to\gamma e^+e^-$. The uncertainties from the tracking and PID of $K^\pm$ and $\pi^0$ reconstruction are obtained by studying a DT control sample $\psi(3770)\to D\bar D$ with hadronic $D$ decays.
The systematic uncertainties from the tracking (PID) efficiencies are assigned as 1.0\% (1.0\%) per $e^\pm$ and 1.0\% (1.0\%) per $K^\pm$, respectively~\cite{BESIII:2021mfl}.
Due to the limited sample size, the systematic uncertainty of the $\eta$ reconstruction is estimated by referring to that of $\pi^0$,
which includes photon efficiency, mass window selection, and invariant mass constraint imposed by kinematic fit. The systematic uncertainty of $\eta$ is taken to be 1.0\% per $\eta$.
The systematic uncertainty associated with the $K^0_S$ reconstruction is assigned to be 1.7\% based on control samples of $\psi(3770)\to D\bar D$.
The systematic uncertainties from the $E_{\rm extra\,\gamma}^{\rm max}$, $N_{\rm extra}^{\rm char}$ and $N_{\rm extra}^{\pi^0}$ requirements are estimated to be 0.9\%, 1.8\%, and 1.8\% for $D^0\to K^-\eta e^+\nu_e$, $D^+\to K^0_S\eta e^+\nu_e$ and $D^+\to \eta \eta e^+\nu_e$ decays, respectively, and are estimated using DT samples of $D^0\to K^-e^+\nu_e$ and $D^+\to K^0_Se^+\nu_e$ decays reconstructed versus the same tags as the default analysis.
The systematic uncertainties from the $M_{P\eta e^{+}}$ requirements are estimated to be 1.7\%, 2.7\%, and 2.7\% for $D^0\to K^-\eta e^+\nu_e$, $D^+\to K^0_S\eta e^+\nu_e$ and $D^+\to \eta \eta e^+\nu_e$, respectively, which are estimated using DT samples of $D^0\to K^-\pi^{0}e^+\nu_e$ and $D^+\to K^0_S\pi^{0}e^+\nu_e$ decays reconstructed versus the same tags as the default analysis.
The contribution of the $M_{K^-\eta}$ requirement to 
the systematic uncertainty is found to be negligible.  

The systematic uncertainty related to the MC generator is estimated using alternative signal MC samples, where the ISGW2 model is replaced by a PHSP model for $D\to {\mathcal R} e^+\nu_e$.
The changes of the signal efficiencies, 1.9\%, 0.1\%, and 0.1\%, are taken as the systematic uncertainties for $D^0\to K^-\eta e^+\nu_e$, $D^+\to K^0_S\eta e^+\nu_e$ and $D^+\to \eta \eta e^+\nu_e$, respectively.
The systematic uncertainty of the MC generator is also considered by varying the mass and width of the assumed resonances according individual PDG values~\cite{pdg2022}. The changes of the signal efficiencies, 7.9\%, 6.0\%, and 6.3\%, are taken as the systematic uncertainties for $D^0\to K^-\eta e^+\nu_e$, $D^+\to K^0_S\eta e^+\nu_e$ and $D^+\to \eta \eta e^+\nu_e$, respectively.
The systematic uncertainties due to unknown resonance are assigned as the differences between the nominal efficiencies and those obtained with ISGW2 or PHSP models, which are 24.4\%, 17.0\%, and 36.4\% for $D^0\to K^-\eta e^+\nu_e$, $D^+\to K^0_S\eta e^+\nu_e$ and $D^+\to \eta \eta e^+\nu_e$, respectively.
Additionally, the uncertainties due to the limited MC sample size, propagated from DT efficiencies, are 0.5\%, 0.4\%, and 0.5\% for $D^0\to K^-\eta e^+\nu_e$, $D^+\to K^0_S\eta e^+\nu_e$ and $D^+\to \eta \eta e^+\nu_e$, respectively.

By summing these uncertainties in quadrature, the total systematic uncertainties associated with the signal efficiencies, $\sigma_{\epsilon}$, are determined to be 26.0\% assumption for $D^0\to K^-\eta e^+\nu_e$, 18.5\% for $D^+\to K^0_S\eta e^+\nu_e$, and 37.2\% for $D^+\to \eta \eta e^+\nu_e$, respectively. These systematic uncertainties are summarized in Table~\ref{table:sys}.

\begin{table*}[htbp]
	\centering
	\caption{Systematic uncertainties in \% for the branching fractions and upper limits, where the dash (-) indicates a systematic effect is not applicable.}
	\label{table:sys}
	\begin{tabular}{lccc}
		\hline\hline
		Source                 & $D^0\to K^-\eta e^+\nu_e$ & $D^+\to K^0_S\eta e^+\nu_e$& $D^+\to \eta \eta e^+\nu_e$ \\ \hline
		$N_{\rm tag}$                                          		 & 0.1           & 0.1             & 0.1             \\
		$K^{-}, e^{+}$ tracking                                                     & 2.0           & 1.0             & 1.0             \\
		$K^{-}, e^{+}$ PID                                                          & 2.0           & 1.0             & 1.0             \\
		$\eta$ reconstruction                                                      & 1.0           & 1.0             & 2.0             \\
		$K_{S}^{0}$ reconstruction                                                 & -           & 1.7             & -             \\
		$E_{\rm extra \gamma}^{\rm max},  N_{\rm extra}^{\rm char},$ and $N_{\rm extra}^{\pi^{0}}$ requirements    & 0.9           & 1.8             & 1.8            \\
		$M_{\eta P e^+}$ requirement											     & 1.7           & 2.7             & 2.7            \\
		Daughter branching fractions         & 0.5           & 0.5             & 0.9            \\
		MC sample size                                                              & 0.5      & 0.4        & 0.5       \\
		MC model                                                                   & 8.1      & 6.0        & 6.3       \\
		Unknown resonance												& 24.4           & 17.0             & 36.4            \\
		$U_{\rm miss}$ fit                                                          & 0.3           & -             & -            \\ \hline
		Total                                                                      & 26.0      & 18.5        & 37.2       \\
		\hline\hline
	\end{tabular}
\end{table*}

The additive systematic uncertainties originate from the fits to the $U_{\rm miss}$ distributions of the semileptonic $D$ decay candidates. They are dominated by the uncertainty from the background shape. This systematic uncertainty is studied by altering the default MC background shape with two methods. First, alternative MC samples are used to determine the background shape, where the relative fractions of the $q\bar{q}$ background is varied within the uncertainties. Second, the BFs of the major $D\bar{D}$ background sources. $D^{0}\to K^{-}\pi^{+}\pi^{0}\pi^{0}$, $D^{0}\to K^{-}\pi^{0}e^{+}\nu_{e}$, $D^{0}\to K^{-}\pi^{+}\eta$, $D^{+}\to K_{S}^{0}\pi^{0}e^{+}\nu_{e}$, $D^{+}\to K_{S}^{0}\pi^{+}\eta$, $D^{+}\to K_{S}^{0}\pi^{+}\pi^{0}\pi^{0}$, $D^{+}\to K_{S}^{0}\pi^{0}e^{+}\nu_{e}$, $D^{+}\to K_{S}^{0}e^{+}\nu_{e}$, and $D^{+}\to K_{S}^{0}\pi^{0}\pi^{+}$ are varied within their uncertainties~\cite{pdg2022}. The signal shape is varied from the MC simulated shape to an analytical shape described by a double-Gaussian function, with the means, widths, and relative areas of the two Gaussian components fixed from a fit to the signal MC sample. The uncertainty associated with the fit to the $U_{\rm miss}$ distribution is estimated to be 0.3\% for the $D^0\to K^-\eta e^+\nu_e$ decay, and is also summarized in Table~\ref{table:sys}. The additive systematic uncertainties for $D^+\to K^0_S\eta e^+\nu_e$ and $D^+\to \eta \eta e^+\nu_e$ are considered in the maximum-likelihood results in next section.

\section{Results}

To take into account the additive systematic uncertainty in the limits, the maximum-likelihood fits are repeated using
different alternative background shapes as mentioned in the previous section and the one resulting in the most conservative upper limit is chosen.
To incorporate the multiplicative systematic uncertainties in the calculation of the upper limits, the likelihood distribution is smeared by a Gaussian function with a mean of zero and a width equal to $\sigma_{\epsilon}$~\cite{stenson2006exact}
\begin{equation}
	L(\mathcal{B})\propto\int_{0}^{1}L\left(\mathcal{B}\frac{\epsilon }{\epsilon_{0}}\right)e^{-\frac{\left(\tfrac{\epsilon}{\epsilon_{0}}-1\right)^2}{2\sigma_{\epsilon}^2}}d\epsilon,
\end{equation}
where $L(\mathcal{B})$ is the likelihood distribution as a function of assumed BFs, $\epsilon$ is the expected efficiency and $\epsilon_{0}$ is the average MC-estimated efficiency.

The red solid curves in Fig.~\ref{fig:prob} show the likelihood distributions incorporating the systematic uncertainties for $D^+\to K_S^0 \eta e^+\nu_e$ and $D^+\to \eta\eta e^+\nu_e$ decays, respectively.
The upper limits on the product of branching fractions of $D^+\to K_S^0 \eta e^+\nu_e$ and $D^+\to \eta\eta e^+\nu_e$ at the 90\% C.L are $2.0\times10^{-4}$ and $1.0\times10^{-4}$, respectively.

Finally, we note that the additive systematic does not affect the significance 
of the $D^0\to K^-\eta e^+\nu_e$ mode within the quoted precision.

\begin{table*}[htbp]
	\centering
	\caption{The signal yields ($N_{\rm sig}^{\rm fit}$), the statistical significances, the signal efficiencies ($\bar{\varepsilon}_{\rm SL}$), the branching fractions ($\mathcal{B}$), the upper limits on the accepted events ($N_{\rm sig}^{\rm up}$), and the upper limits of the branching fractions at the 90\% confidence level, ($\mathcal{B}^{\rm up}$) for $D^0\to K^-\eta e^+\nu_e$, $D^+\to K_S^0 \eta e^+\nu_e$ and $D^+\to \eta\eta e^+\nu_e$ decays.}
	\label{table:bfalldata}
	\begin{tabular}{lcccccc}
		\hline\hline
		Decay                        & $N_{\rm sig}^{\rm fit}$ & Significance~($\sigma$) & $N_{\rm sig}^{\rm up}$ & $\bar{\varepsilon}_{\rm SL}~(\%)$ & $\mathcal{B}~(\times10^{-4})$ & $\mathcal{B}^{\rm up}~(\times10^{-4})$ \\ \hline
		$D^0\to K^-\eta e^+\nu_e$    & $11.1_{-3.8}^{+4.5}$    & 3.3                     & ...                    & $5.29 \pm 0.05$                   & $0.84_{-0.29}^{+0.34}\pm0.22$ & ...                                    \\
		$D^+\to K_S^0 \eta e^+\nu_e$ & $8.4_{-3.9}^{+4.7}$     & 1.9                     & 17.8                   & $7.79 \pm 0.05$                   & ...                           & $<2.0$                                 \\
		$D^+\to \eta\eta e^+\nu_e$   & $0.3_{-1.2}^{+1.9}$     & 0.0                     & 4.8                    & $7.52 \pm 0.05$                   & ...                           & $<1.0$                                 \\ \hline\hline
	\end{tabular}
\end{table*}

\section{Summary}

Based on a $7.93~\mathrm{fb}^{-1}$ sample of $e^+e^-$ collision data taken at $\sqrt{s}=3.773$~GeV with the BESIII detector, the semileptonic decays $D^0\to K^-\eta e^+\nu_e$, $D^+\to K_S^0 \eta e^+\nu_e$ and $D^+\to \eta\eta e^+\nu_e$ have been investigated for the first time.
Evidence for $D^0\to K^-\eta e^+\nu_e$ is found with a significance of $3.3\sigma$. The branching fraction of $D^0\to K^-\eta e^+\nu_e$ is determined to be $(0.84_{-0.34}^{+0.29}\pm0.22)\times 10^{-4}$.
No significant signals are found for $D^+\to K_S^0 \eta e^+\nu_e$ and $D^+\to \eta\eta e^+\nu_e$ and the upper limits on their branching fractions are set at the 90\% C.L.
With the currently available data sample, the obtained upper limits are comparable with the theoretical calculations. The detailed results are presented in Table~\ref{table:bfalldata}.

\section{Acknowledgement}

The BESIII Collaboration thanks the staff of BEPCII and the IHEP computing center for their strong support. This work is supported in part by National Key R\&D Program of China under Contracts Nos. 2023YFA1606000, 2020YFA0406400, 2020YFA0406300; National Natural Science Foundation of China (NSFC) under Contracts Nos. 11635010, 11735014, 11935015, 11935016, 11935018, 12025502, 12035009, 12035013, 12061131003, 12192260, 12192261, 12192262, 12192263, 12192264, 12192265, 12221005, 12225509, 12235017, 12361141819; the Chinese Academy of Sciences (CAS) Large-Scale Scientific Facility Program; the CAS Center for Excellence in Particle Physics (CCEPP); Joint Large-Scale Scientific Facility Funds of the NSFC and CAS under Contract No. U1832207; 100 Talents Program of CAS; The Institute of Nuclear and Particle Physics (INPAC) and Shanghai Key Laboratory for Particle Physics and Cosmology; German Research Foundation DFG under Contracts Nos. 455635585, FOR5327, GRK 2149; Istituto Nazionale di Fisica Nucleare, Italy; Ministry of Development of Turkey under Contract No. DPT2006K-120470; National Research Foundation of Korea under Contract No. NRF-2022R1A2C1092335; National Science and Technology fund of Mongolia; National Science Research and Innovation Fund (NSRF) via the Program Management Unit for Human Resources \& Institutional Development, Research and Innovation of Thailand under Contract No. B16F640076; Polish National Science Centre under Contract No. 2019/35/O/ST2/02907; The Swedish Research Council; U. S. Department of Energy under Contract No. DE-FG02-05ER41374.



\end{document}